\documentclass[reprint,showpacs,preprintnumbers,amsmath,amssymb,aps]{revtex4-1}

\usepackage{graphicx}
\usepackage{dcolumn}
\usepackage{bm}
\usepackage{epsfig}
\usepackage{subfigure}

\begin{document}


\title{Anti-phase synchronization of phase-reduced oscillators using open-loop control}

\author{Dionisis Stefanatos}
\email{dionisis@seas.wustl.edu}
\author{Jr-Shin Li}
\email{jsli@seas.wustl.edu}
\affiliation{Department of Electrical and Systems Engineering, Washington University, St. Louis, MO 63130, USA}

\date{\today}

\begin{abstract}
In this letter, we present an elegant method to build and maintain an anti-phase configuration of two nonlinear oscillators with different natural frequencies and dynamics described by the sinusoidal phase-reduced model. The anti-phase synchronization is achieved using a common input that couples the oscillators and consists of a sequence of square pulses of appropriate amplitude and duration. This example provides a proof of principle that open-loop control can be used to create desired synchronization patterns for nonlinear oscillators, when feedback is expensive or impossible to obtain.
\end{abstract}

\maketitle


Natural and artificial systems that consist of collections of isolated or interacting nonlinear oscillators are reaching complexity levels that are beyond human understanding. The normal operation of these complex systems
often requires the formation of certain synchronization structures. The synchronization of oscillating systems is a fundamental and extensively studied phenomenon in natural sciences and engineering \cite{Pikovsky01,Strogatz94}. Examples include circadian rhythms \cite{Winfree80}, neural circuitry in the brain \cite{Uhlhaas06}, pacemaker cells of the heart \cite{Jalife84}, insulin-secreting cells of the pancreas \cite{Sherman88}, chemical oscillations \cite{Kuramoto84}, semiconductor lasers in physics \cite{Fischer00}, and vibrating systems in mechanical engineering \cite{Blekhman88}. The above systems are often extremely large in scale, which poses serious theoretical and computational challenges to efforts to model, guide, or optimize
them. Deriving control signals that can drive complex systems to desired synchronization configurations is of utmost theoretical and practical importance \cite{Kiss07}. A premier example comes from the area of neuroscience, where devising low-power external stimuli that synchronize or desynchronize a network of coupled or uncoupled neurons is imperative for wide-ranging applications, from neurological treatment of Parkinson's disease and epilepsy \cite{Ashwin92,Benabid91,Schiff94} to the design of neurocomputers \cite{Hoppensteadt99,Hoppensteadt00}.

Mathematical devices are required for describing the complex dynamics of oscillating systems
in a manner that is both tractable and flexible in design. A promising approach
to construct simplified yet accurate models that capture the essential
overall system properties is through the use of phase model reduction.
Underlying this method is the fact that an oscillating system
with a stable periodic orbit, described by a system of first order ordinary differential equations, can be
reduced, under certain circumstances, to a dynamic equation in a single variable, which represents the
phase $\theta$ of the system on its limit cycle \cite{Winfree80, Kuramoto84}
\begin{equation}
\label{phase_model}
\dot{\theta}=f(\theta)+u(t)Z(\theta),
\end{equation}
where $u(t)$ is some external input, for example a current, $f(\theta)$ gives the baseline dynamics, which is present even when $u=0$, and $Z(\theta)$ describes the phase sensitivity to the stimulus and is called the phase response curve (PRC).
For example, for the prototype mathematical neuron model, the famous Hodgkin-Huxley model which describes the propagation of action potentials along the giant squid axon and is a complex system
of four highly nonlinear differential equations \cite{Hodgkin52}, the corresponding phase-reduced model can be computed \cite{Moehlis06} using widely available software \cite{Ermentrout02}.

Phase models have been very effectively used in
theoretical, numerical, and more recently, experimental studies to analyze the collective behavior
of networks of oscillators \cite{Preyer05, Acebron05, Kiss05, Netoff05}.
Specifically, these models have been
used to investigate either the patterns of synchrony that result
from the type and architecture of coupling \cite{Cohen82,Kopell90,Ashwin92,Hansel93,Orosz09},
or the response of the system to external stimuli \cite{Hoppensteadt99, Brown04}.
Motivated by studies in control and systems theory which show that \emph{feedback} is essential for investigating properties of complex dynamical systems such as self-organization and stability \cite{Zecevic10}, various feedback approaches, based on phase models, have recently been developed for the efficient control of synchronization patterns in oscillator assemblies \cite{Kiss07,Zhai08,Rusin10}. Although these synchronization engineering methods are effective for the synthesis of subtle dynamical patterns such as itinerant cluster dynamics, in many emerging applications involving the control of large-scale complex systems, state feedback maybe difficult, impossible, or expensive to obtain, or the types of feedback laws that can be used are severely restricted due to the complexity of system dynamics. As a result, the development of \emph{open-loop} controls for the design of phase patterns for oscillator ensembles is compelling.

Recently, geometric control theory \cite{Jurdjevic96} has been employed to study controllability of a network of oscillators with different natural oscillation frequencies \cite{Li10}. Controllability guarantees the existence of open-loop inputs that can drive a system of oscillators to any desired synchronization pattern. The idea to use an open-loop control to achieve the target synchronization structure is still useful even when the system is not fully controllable but the final state is reachable \cite{Becker10}.

In this letter, we demonstrate the potential of open-loop control to engineer patterns of synchronization in collections of nonlinear oscillators, using a simple yet important system. Specifically, we show how an input sequence composed of surprisingly simple square pulses can be used to build and maintain a $\pi$ phase difference between two nonlinear oscillators with dynamics described by a sinusoidal phase-reduced model, with $f$ constant and $Z(\theta)=\sin{\theta}$ in (\ref{phase_model}). This model is an approximation
to the FitzHugh-Nagumo model, a two-dimensional simplification of the Hodgkin-Huxley neuron model, near the Hopf bifurcation point \cite{Keener01}. If Parkinson's disease is considered as the synchronized response of an
assembly of neurons, described by phase models, then it is desirable to bring these oscillators out
of phase with an open-loop signal. This observation highlights the importance of establishing an anti-phase configuration for a system of two phase-reduced oscillators.

In the following we consider a pair of oscillators described by the sinusoidal phase-reduced model
\begin{eqnarray}
\label{system1}\dot{\theta}_1 & = & \omega_1+u\sin{\theta_1},\\
\label{system2}\dot{\theta}_2 & = & \omega_2+u\sin{\theta_2},
\end{eqnarray}
where $\omega_2>\omega_1$ but close to each other. The common input $u(t)$ couples the two oscillators and we suppose that it is bounded by $|u|<\omega_1$. 
We assume that a well-defined initial state can be established by a classical phase resetting method, where a large pulse is applied to the oscillators and brings them to the same initial phase, independent of the phases they had before \cite{Winfree80}. The goal is, starting from $\theta_1(0)=\theta_2(0)=0$, to build a $\pi$ phase difference between the two oscillators and then maintain this anti-phase configuration. We describe how this can be done efficiently using simple square pulses. The approach is different from that in \cite{Hoppensteadt99}, where the phase difference is built asymptotically in time.

First we show how this anti-phase configuration can be produced by applying a constant signal $u=-M$ of appropriate magnitude for a duration $T_M$. If we require at time $t=T_M$ that
\begin{equation*}
\theta_1(T_M)=(2n-1)\pi,\quad\theta_2(T_M)=2n\pi
\end{equation*}
for some positive integer $n$, then, using (\ref{system1}) and (\ref{system2}), we obtain
\begin{equation*}
\int_{0}^{(2n-1)\pi}\frac{d\theta_1}{\omega_1-M\sin{\theta_1}}=\int_{0}^{2n\pi}\frac{d\theta_2}{\omega_2-M\sin{\theta_2}}=T_M,
\end{equation*}
or, after integrating,
\begin{eqnarray}
\frac{2}{\sqrt{\omega_1^2-M^2}}\left[\frac{(2n-1)\pi}{2}+\tan^{-1}\left(\frac{M}{\sqrt{\omega_1^2-M^2}}\right)\right]=&\nonumber\\\label{transM}\frac{2n\pi}{\sqrt{\omega_2^2-M^2}}.\quad
\end{eqnarray}

Let $L(M)$ and $R(M)$ denote the left and right hand sides of (\ref{transM}), respectively, as functions of the amplitude $M$. It is $L(0)=(2n-1)\pi/\omega_1, R(0)=2n\pi/\omega_2$ and $L(\omega_1)\rightarrow\infty, R(\omega_1)=2n\pi/\sqrt{\omega_2^2-\omega_1^2}$. If
\begin{equation}
\label{ratio1}
\frac{\omega_2}{\omega_1}<\frac{2n}{2n-1},
\end{equation}
then $L(0)<R(0)$, while it is always $L(\omega_1)>R(\omega_1)$. Thus, for $\omega_1$ sufficiently close to $\omega_2$, the transcendental equation (\ref{transM}) has a solution $M\in(0,\omega_1)$.
The duration $T_M$ of the constant signal $u=-M$ is given by
\begin{equation*}
\label{TM}
T_M=R(M)=\frac{2n\pi}{\sqrt{\omega_2^2-M^2}}.
\end{equation*}
Note that in the absence of any input, i.e. when $u=0$, a time $T_0=\pi/(\omega_2-\omega_1)$ is necessary to build a $\pi$ difference between the oscillators. Also, observe that for given frequencies $\omega_1$ and $\omega_2$, (\ref{ratio1}) can be used to calculate the allowed values of $n$. As $n$ increases, we expect that a signal with lower amplitude $M$ and larger duration $T_M$ is required to create the $\pi$ phase difference.

\begin{figure*}[t!]
 \centering
		\begin{tabular}{cc}
     	\subfigure[$\ $Common input, $n=1$]{
	            \label{fig:control1}
	            \includegraphics[width=.35\linewidth]{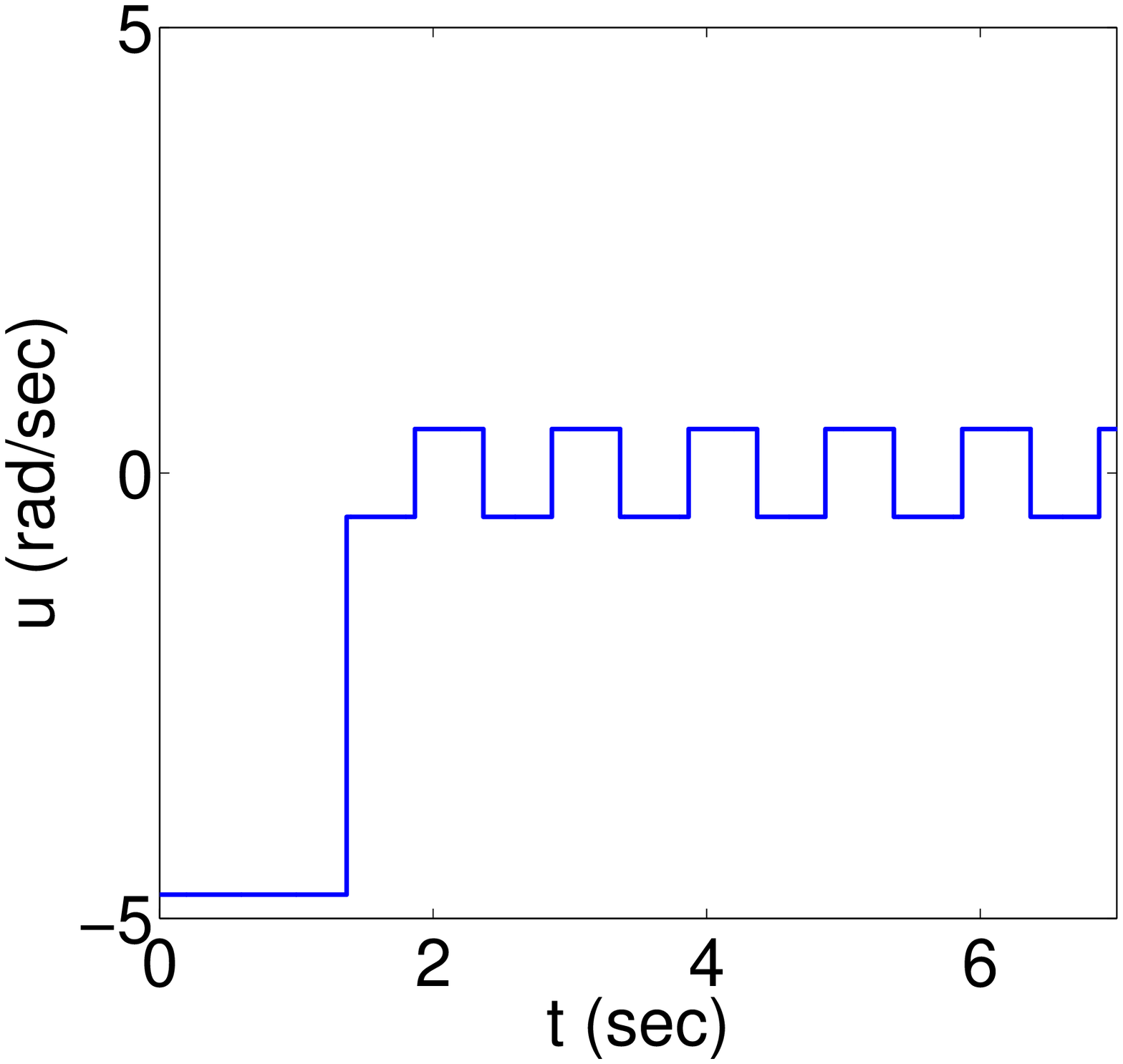}} &
        \subfigure[$\ $Phase difference, $n=1$]{
	            \label{fig:difference1}
	            \includegraphics[width=.35\linewidth]{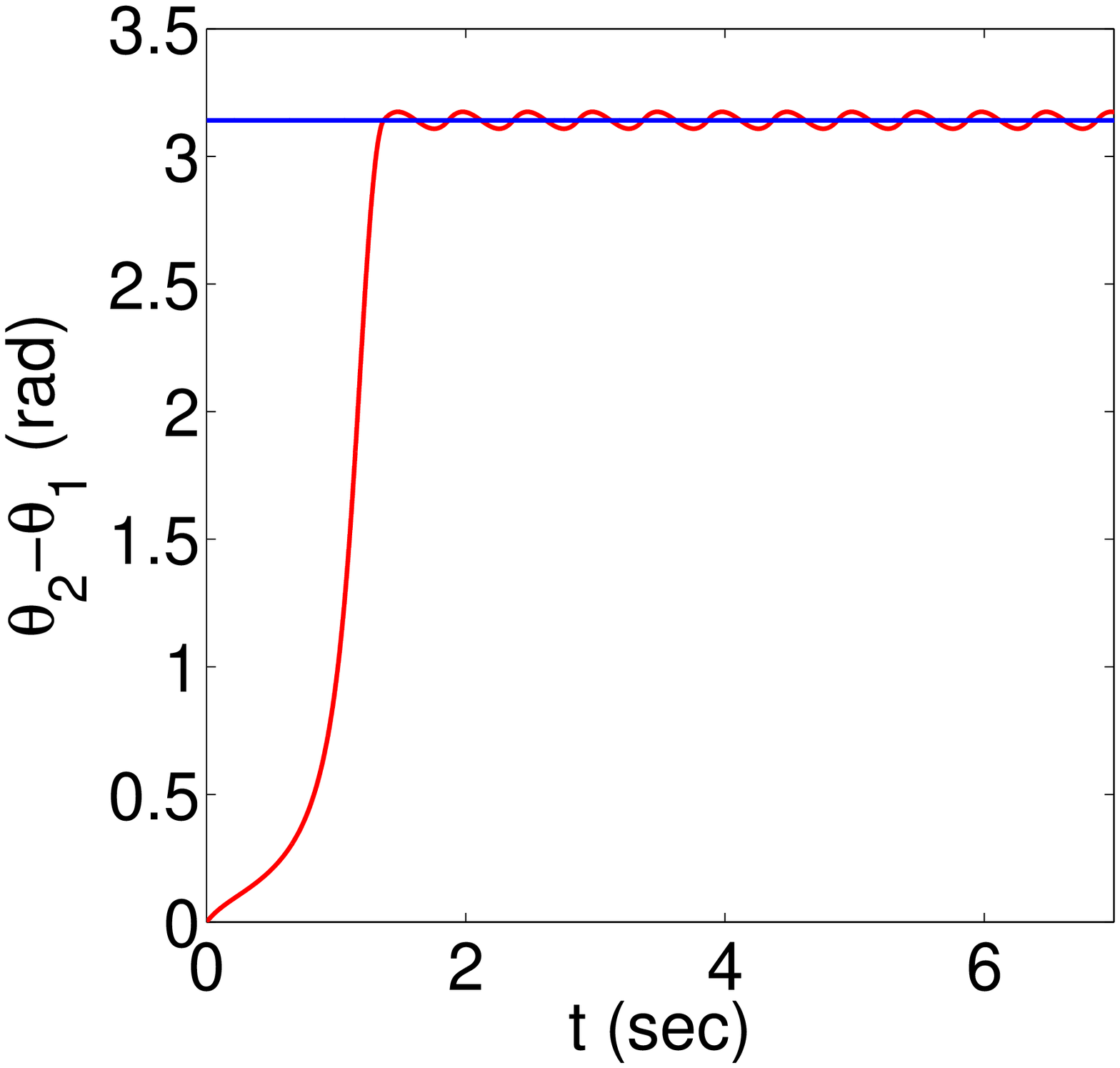}} \\
	    \subfigure[$\ $Common input, $n=3$]{
	            \label{fig:control3}
	            \includegraphics[width=.35\linewidth]{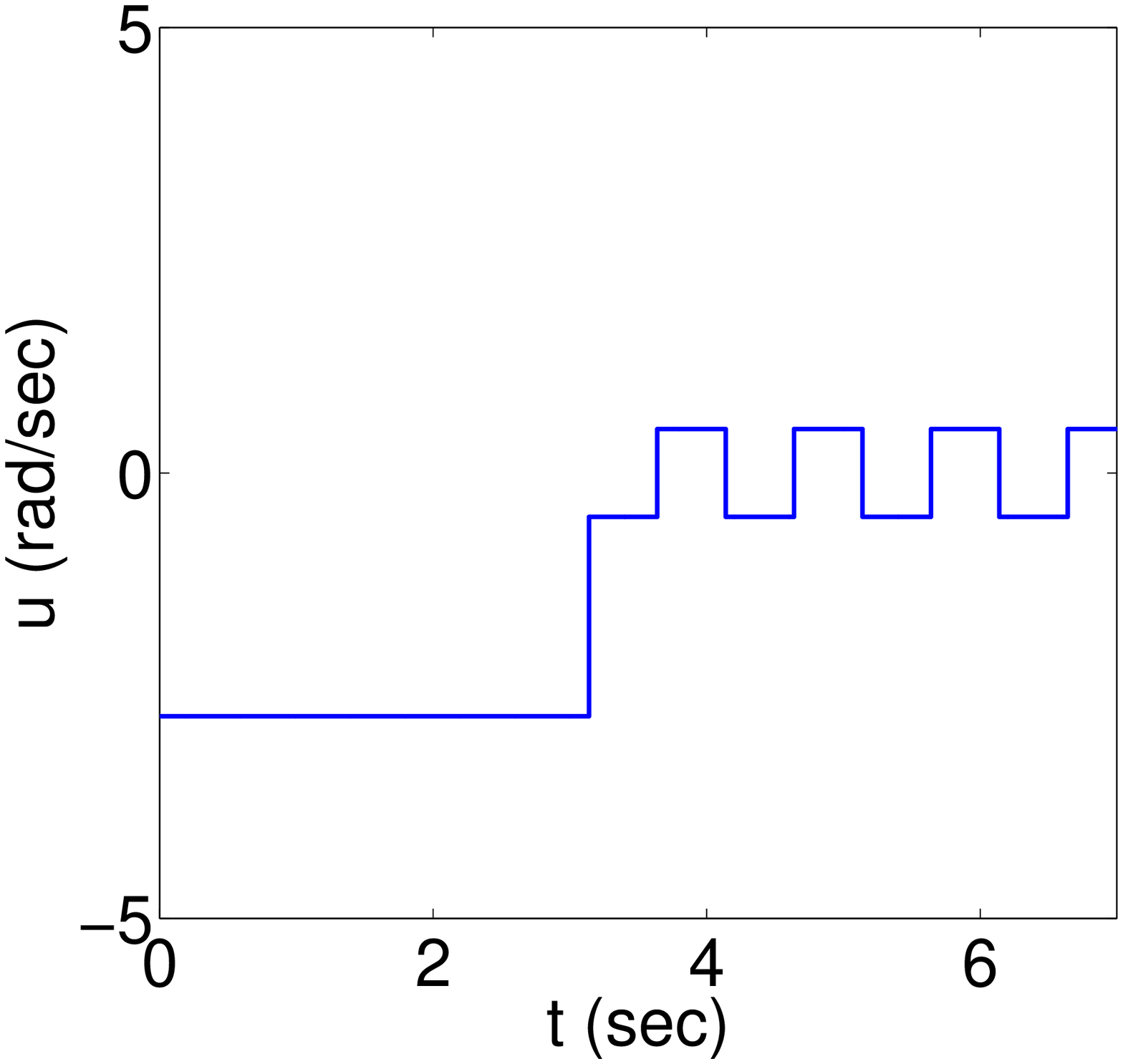}} &
        \subfigure[$\ $Phase difference, $n=3$]{
	            \label{fig:difference3}
	            \includegraphics[width=.35\linewidth]{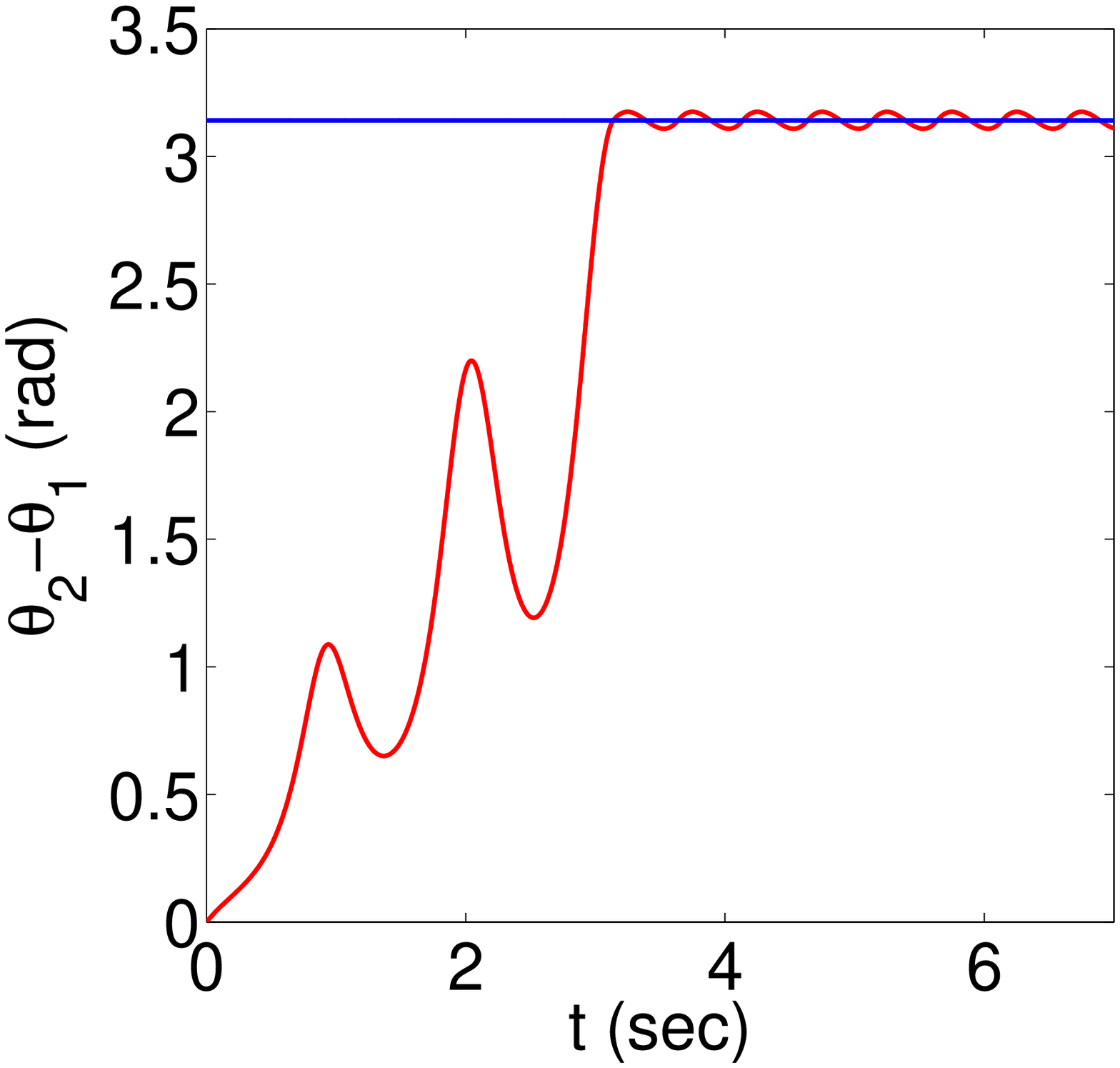}} \\
        \subfigure[$\ $Common input, $n=5$]{
	            \label{fig:control5}
	            \includegraphics[width=.35\linewidth]{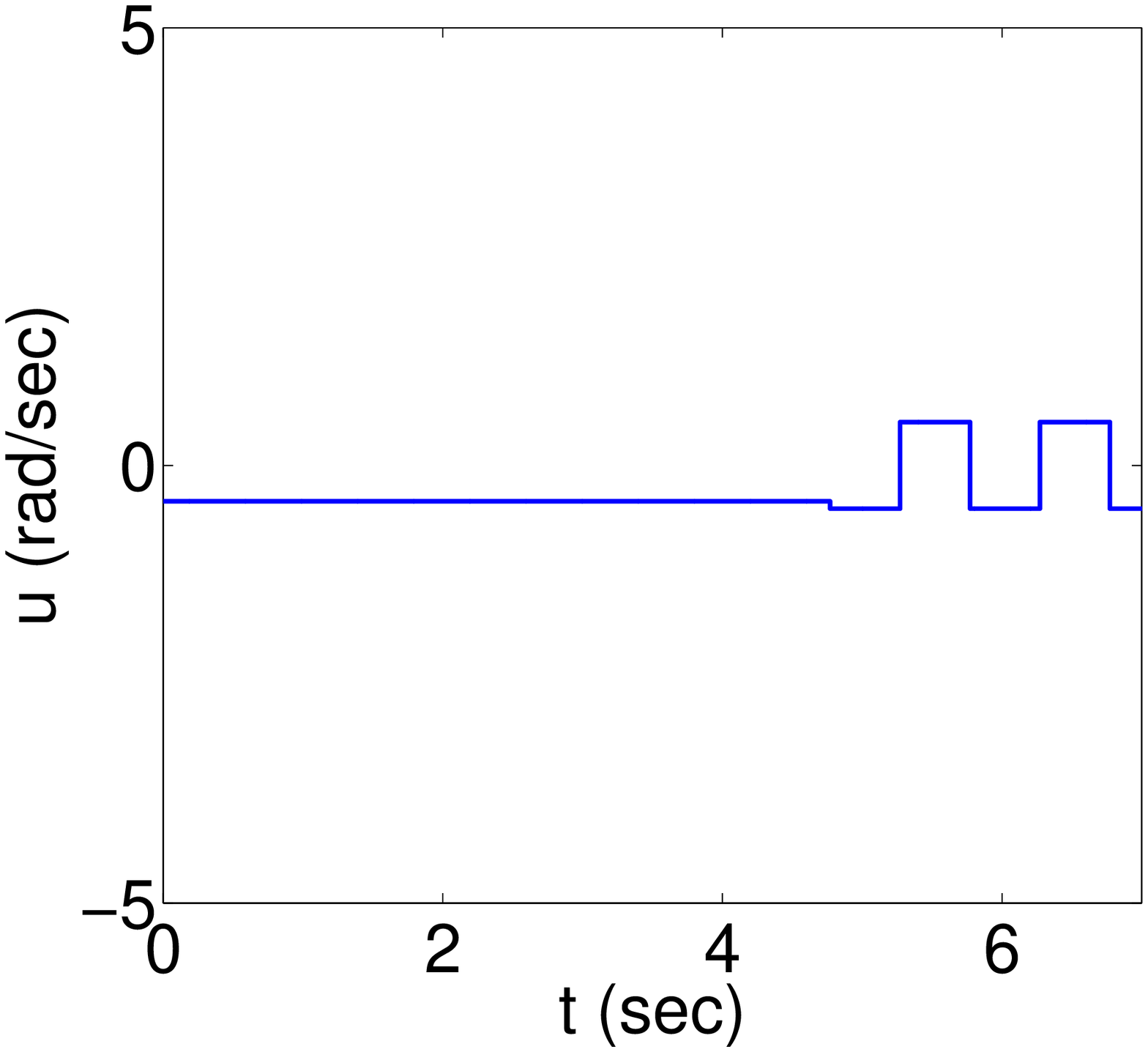}} &
        \subfigure[$\ $Phase difference, $n=5$]{
	            \label{fig:difference5}
	            \includegraphics[width=.35\linewidth]{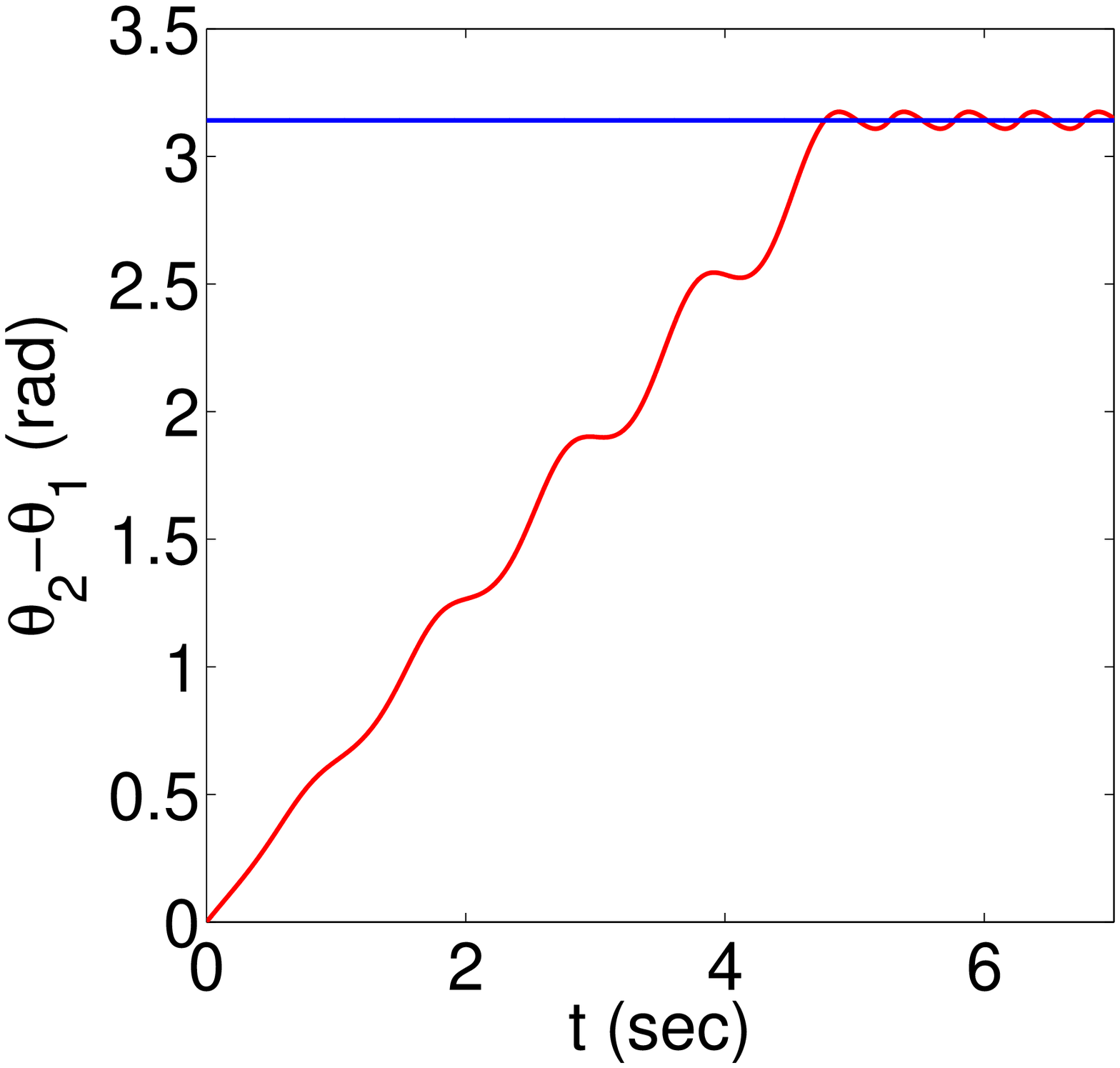}}
		\end{tabular}
\caption{(Color online) For $\omega_1=1.9\pi$ rad/sec and $\omega_2=2.1\pi$ rad/sec we plot the input function (a, c, e) and the phase difference $\theta_2-\theta_1$ (b, d, f), for $n=1,3,5$. Observe that for larger $n$, a smaller amplitude is necessary to build the $\pi$ phase difference. After this difference has been built, a simple square wave is sufficient to maintain it, the same for all $n$. The small periodic deviation from $\pi$ is expected since the method is designed to preserve the anti-phase configuration at certain instants separated by half period of the square wave.}
 \label{fig:oneswitching}
\end{figure*}

Once the anti-phase configuration has been set up, we can maintain it by appropriately choosing the control $u$. The idea is that when $\theta_2-\theta_1=\pi$, then $\sin{\theta_1}$ and $\sin{\theta_2}$ have always opposite signs, so we can use as common input a square wave with appropriate amplitude $m$ and period $T_m$
\begin{equation*}
u(t)=(-1)^km,\quad T_M+\frac{k-1}{2}T_m< t\leq T_M+\frac{k}{2}T_m,
\end{equation*}
$k=1,2,\ldots$, to accelerate the slow oscillator and decelerate the fast one.

The control amplitude $m$ is chosen to assure that phase difference is preserved at the switching instants of the square wave
\begin{equation*}
\theta_2\left (T_M+\frac{k}{2}T_m\right)-\theta_1\left (T_M+\frac{k}{2}T_m\right)=\pi.
\end{equation*}
This implies that the modified half-periods
of the two oscillators for $t>T_M$, after the application of the square wave, are the same and equal to the half-period $T_m/2$ of this wave. If, without loss of generality, we use the first half-period $T_M<t\leq T_M+T_m/2$ after the onset of the square wave, we obtain the condition
\begin{equation}
\label{condition1}
\int_{(2n-1)\pi}^{2n\pi}\frac{d\theta_1}{\omega_1-m\sin{\theta_1}}=\int_{2n\pi}^{(2n+1)\pi}\frac{d\theta_2}{\omega_2-m\sin{\theta_2}}=\frac{T_m}{2},
\end{equation}
or, after the change of variables $\phi_1=\theta_1-(2n-1)\pi,
\phi_2=\theta_2-2n\pi$,
\begin{equation}
\label{condition2}
\int_{0}^{\pi}\frac{d\phi_1}{\omega_1+m\sin{\phi_1}}=\int_{0}^{\pi}\frac{d\phi_2}{\omega_2-m\sin{\phi_2}}.
\end{equation}
Observe that different integration intervals in (\ref{condition1}), due to the already built $\pi$ difference, led to a different control sign in the denominator of the integrands in (\ref{condition2}), increasing thus the period of the fast oscillator and decreasing that of the slower.
Integrating (\ref{condition2}) yields
\begin{eqnarray}
\frac{2}{\sqrt{\omega_1^2-m^2}}\left[\frac{\pi}{2}-\tan^{-1}\left(\frac{m}{\sqrt{\omega_1^2-m^2}}\right)\right]&=\nonumber\\
\label{transm}\frac{2}{\sqrt{\omega_2^2-m^2}}\left[\frac{\pi}{2}+\tan^{-1}\left(\frac{m}{\sqrt{\omega_2^2-m^2}}\right)\right].&
\end{eqnarray}


If we use $L(m)$ and $R(m)$ to denote the left and right hand sides of (\ref{transm}), as functions of the amplitude $m$, then $L(0)=\pi/\omega_1, R(0)=\pi/\omega_2$ and $L(\omega_1)\rightarrow 2/\omega_1, R(\omega_1)>\pi/\sqrt{\omega_2^2-\omega_1^2}$. Since $\omega_2>\omega_1$, it is $L(0)>R(0)$. The requirement $L(\omega_1)<R(\omega_1)$ is satisfied when $2/\omega_1<\pi/\sqrt{\omega_2^2-\omega_1^2}$, which leads to the condition
\begin{equation}
\label{ratio2}
\frac{\omega_2}{\omega_1}<\sqrt{1+\left(\frac{\pi}{2}\right)^2}.
\end{equation}
When (\ref{ratio2}) holds, the transcendental equation (\ref{transm}) has a solution $m\in(0,\omega_1)$.
The modified period is
\begin{equation*}
\label{Tm}
T_m=2L(m)=\frac{4}{\sqrt{\omega_1^2-m^2}}\tan^{-1}\left(\frac{\sqrt{\omega_1^2-m^2}}{m}\right).
\end{equation*}

For $\omega_1=1.9\pi$ rad/sec and $\omega_2=2.1\pi$ rad/sec, condition (\ref{ratio2}) is satisfied and from (\ref{ratio1}) we find $n\leq 5$.
In Fig. \ref{fig:oneswitching} we plot the input sequence (left column) and the resulting phase difference (right column), for $n=1,3,5$. Observe that for increasing $n$ the necessary amplitude $M$ to build the desired phase difference is smaller but the corresponding time $T_M$ is larger. The maintenance signal is the same in all cases. Note that after the establishment of the phase difference there is a small periodic deviation from $\pi$, but this is expected since the method is designed to maintain the anti-phase configuration at certain time instants separated by $\Delta t=T_m/2$.


We have demonstrated that open-loop control can be used to design patterns of synchrony for nonlinear oscillators, by
presenting a simple and elegant method which can achieve and maintain a $\pi$ phase difference between two phase-reduced oscillators using surprisingly simple square pulses. This approach can find applications across various disciplines in natural sciences where synchronization plays an important role, to situations where feedback is expensive or impossible to obtain. 

This work was supported by the Air Force Office of Scientific Research under grant \#FA9550-10-1-0146.




\end{document}